\documentclass[10pt]{iopart}  
\usepackage{graphicx}
\usepackage{braket}
\usepackage{algpseudocode}

\newcommand{\Eq}[1]{\textup(\ref{eq:#1})} 
\newcommand{\operatorname}[1]{\textrm{#1}}  
\newcommand{\text}[1]{\textrm{#1}}  
\newcommand{\lessapprox}{{<\approx}}  
\newcommand{\url}[1]{\texttt{#1}}  
\newcommand{\dd}{\, \rmd} 
\newcommand{\ii}{\rmi} 
\newcommand{\Op}[1]{\ensuremath{\mathsf{\hat{#1}}}}
\newcommand{\Abs}[1]{\left|#1\right|}

\newcommand{\prob}{\operatorname{p}}

\newcommand{\effective}{\operatorname{eff}}

\newcommand{\Norm}[1]{\left|\!\left|#1\right|\!\right|}
\newcommand{\KetBra}[2]{\ket{#1}\!\bra{#2}}
\newcommand{\DynMap}{\mathcal{E}}
\newcommand{\Lindbladian}{\mathcal{L}}

\newcommand{\OpPiGround}{\Op{\Pi}_{\text{g}}}
\newcommand{\OpPiExc}{\Op{\Pi}_{\text{e}}}
\newcommand{\OpSigma}[2]{\Op{\sigma}_{\text{#1}, \text{#2}}}
\newcommand{\KetG}{\ket{\text{g}}}
\newcommand{\KetE}{\ket{\text{e}}}
\newcommand{\KetR}{\ket{\text{r}}}
\newcommand{\Avg}[1]{\langle{#1}\rangle}
\newcommand{\LBra}[1]{\langle\kern-0.2em\langle{#1}\vert}
\newcommand{\LKet}[1]{\vert{#1}\rangle\kern-0.2em\rangle}
\newcommand{\biggLBra}[1]{\bigg\langle\kern-0.5em\bigg\langle{#1}\bigg\vert}
\newcommand{\biggLKet}[1]{\bigg\vert{#1}\bigg\rangle\kern-0.5em\bigg\rangle}
\newcommand{\LBraket}[2]{\langle\kern-0.2em\langle{#1}\vert{#2}\rangle\kern-0.2em\rangle}
\newcommand{\bigLBraket}[2]{\big\langle\kern-0.25em\big\langle{#1}\big\vert{#2}\big\rangle\kern-0.25em\big\rangle}

\renewcommand{\Im}{\operatorname{Im}}

\usepackage{enumitem}
\setitemize{noitemsep,topsep=0pt,parsep=0pt,partopsep=0pt,label={--}}

\newcommand{\Stanford}{Edward L. Ginzton Laboratory, Stanford University,
Stanford, CA 94305, USA.}
\newcommand{\ARL}{U.S. Army Research Laboratory, Computational and Information
Sciences Directorate, Adelphi, MD 20783, USA.}
\newcommand{\UBoston}{Department of Physics, University of Massachusetts at
Boston, Boston, MA 02125, USA}
\newcommand{\Hearne}{Hearne Institute for Theoretical Physics, Louisiana State
University, Baton Rouge, LA 70803, USA}

\makeatletter
\def\Fig{\@ifnextchar[{\@FigWith}{\@FigWithout}}
\def\@FigWith[#1]#2{Fig.~\ref{fig:#2}\,(#1)}
\def\@FigWithout#1{Fig.~\ref{fig:#1}}
\def\Figure{\@ifnextchar[{\@FigureWith}{\@FigureWithout}}
\def\@FigureWith[#1]#2{Figure~\ref{fig:#2}\,(#1)}
\def\@FigureWithout#1{Figure~\ref{fig:#1}}
\makeatother

\begin{document}

\title[Efficient optimization using quantum trajectories]
{Efficient optimization of state preparation in quantum networks using quantum trajectories}


\author{Michael H Goerz$^{1,2}$, Kurt Jacobs$^{2,3,4}$}  
\ead{goerz@stanford.edu}  
\address{$^1$\Stanford} \address{$^2$\ARL}  
\address{$^3$\UBoston} \address{$^4$\Hearne}  

\date{\today}

\begin{abstract}
The wave-function Monte-Carlo method, also referred to as the use of
``quantum-jump trajectories'', allows efficient simulation of open systems by
independently tracking the evolution of many pure-state ``trajectories''.
This method is ideally suited to simulation by modern, highly parallel
computers.
Here we show that Krotov's method of numerical optimal control, unlike others,
can be modified in a simple way so that it becomes fully parallel in the pure
states without losing its effectiveness.
This provides a highly efficient method for finding optimal control protocols
for open quantum systems and networks.
We apply this method to the problem of generating entangled states in a network
consisting of systems coupled in a unidirectional chain.
We show that due to the existence of a dark-state subspace in the network,
nearly-optimal control protocols can be found for this problem by using only a
single pure-state trajectory in the optimization, further increasing the
efficiency.

\end{abstract}

\maketitle

\section{Introduction}

The control of simple quantum systems is central to the future realization of
quantum technologies~\cite{Kohout02,Giovannetti11,Stephens14,Roy17}.
Such systems are usually open, meaning that they experience a significant source
of noise from their environments~\cite{Breuer07,Jacobs14}.
Systems that form the nodes in a quantum network, in which the nodes are
connected together via traveling-wave fields are also necessarily open due to
the interaction with the fields~\cite{Jacobs14,Combes17,gardiner85, gardiner04}.
At the simplest level, open quantum systems and networks can be described by a
first-order differential equation for the density matrix called a master
equation.
A useful feature of master equations is that the evolving density matrix for the
system or systems can alternatively be written as the average over an ensemble
of pure states in which the evolution of each state is given by an equation with
a random (stochastic) component~\cite{Carm93,Diosi86,WisemanLinQ,Wiseman93,
Jacobs14}.
For a $d$-dimensional system the representation of the density matrix scales as
$d^2$ while that of a pure state scales only as $d$.
For this reason the pure-state ensemble approach, or ``quantum trajectory''
approach, can provide a significant reduction in numerical overhead.
Here we use the trajectory method in which the stochastic element is a series of
``jumps'' at random times,  often referred to as the ``quantum Monte Carlo'' or
``quantum-jump trajectories'' method~\cite{Hegerfeldt92,DumPRA1992,
MolmerJOSA1993}.

The ability to vary one or more parameters of the Hamiltonian of a system over
time is a powerful tool for controlling the system.
The problem of determining the time-dependence, or ``pulse shapes'' required to
realize a given evolution is highly non-trivial, however, and as a result
numerical search methods, often referred to as ``optimal control'' methods, are
very useful for this purpose~\cite{GlaserEPJD2015,KochJPCM2016}.
To perform a numerical search requires simulating the evolution of a system or
network many times under differing pulse shapes to sufficiently explore the
``control space'' of possible pulses.
The use of quantum trajectories for such simulations, especially as the size of
a network increases, is thus desirable.

Here we examine the form that gradient-based numerical search methods take when
written in terms of the ensemble of pure state evolutions (the ``quantum
trajectories'') that simulate the master equation.
We show that unlike other methods, Krotov's method can be formulated in such a
way that updates to the pulse shapes are obtained either from independent
trajectories, or by quadratic cross-referencing of trajectories.
We then show via a numerical example that Krotov's method remains effective even
for a small number of trajectories, including a single trajectory.

In the second part of this work we apply the above ``Krotov trajectory method''
to a problem of preparing a class of entangled states in a simple network.
This network consists of a ``chain'' of systems connected via a single
unidirectional field.
Each system, or ``node'' in the network is a cavity containing a system with two
stable levels, and for which the interaction with the cavity mode can be turned
on and off via a local control field.
One way in which these nodes can be realized is by having a three-level atom in
each cavity, in which the three levels are in a lambda configuration.
The two ground states provide the two stable levels, and the interaction between
these two levels and the cavity mode is mediated by an off-resonant coupling to
the excited state.
Our problem may be thought of as a generalization of the configuration used by
Cirac~\textit{et al.\ }in their seminal work in which they showed how the state
of a two-level system in a cavity could be transferred to a second two-level
system in a second cavity via a unidirectional field~\cite{CiracPRL1997}.
They found that, so long as the coupling of each cavity to the traveling-wave
field was the only source of dissipation, the coupling between the two-level
systems and their respective cavities could be controlled in such a way that the
excitation stored in the first two-level system could be emitted into the field,
enter the second cavity, and be completely re-absorbed by the second two-level
system, without being lost from the second cavity to the traveling field (that
is, lost out of the ``open end" of the field that travels away from the
cavities). Strictly speaking, to effect perfect transfer in this way requires an
infinite time, since the transfer of the excitation to the field is an
exponential decay, but near-unity fidelity can be obtained within a finite
multiple of the decay time.
One way to understand the protocol of Cirac~\textit{et al.\ }is to say that
there is a subspace of joint states of the two nodes that are ``dark'' in the
sense that no photons will be emitted from the systems when in this
subspace~\cite{Kraus08, Clark03, Cho11, Stannigel12}.
The protocol is able to realize the transfer by evolving exclusively within this
subspace.
Protocols that employ the transfer method proposed by~Cirac~\textit{et al.\
}have been realized in a number of experiments~\cite{Ritter12, Yin13,
Wenner2014, Axline2017, Kurpiers2017}.

Here we consider the preparation of dark states of $N$ nodes in which a single
excitation, originally prepared in the first node, is shared equally between all
the nodes.
We find that by controlling only the local coupling between the nodes and the
field link it is possible to prepare these multi-partite entangled states while
remaining within the dark-state subspace, thus realizing essentially perfect
preparation (up to the infidelity imposed by the finite duration of the protocol
and any sources of loss and decoherence additional to the field link). We also
find that because the optimal protocols are able to enter the dark-state
subspace quickly on the timescale of the transfer, the majority of the numerical
search can be performed with a single pure-state trajectory, thus greatly
increasing the efficiency.

Our trajectory-based optimization method is applicable to essentially any
control task in which the effect of the environment can be written as an average
over pure-state trajectories.
This is true for all Markovian master equations that have the Lindblad form, and
a few non-Markovian master equations~\cite{Breuer99, Strunz99a, Strunz99b,
Gambetta02b, Gambetta04, Piilo08, Wiseman08b}.
Our method is especially useful for situations in which there are dark-state
spaces in which the effect of the bath vanishes.
This can be true for quantum networks, as in the example we explore here.
Another scenario in which this holds is that of the storage and retrieval of a
photon wave-packet from a collective dark state of an atomic
ensemble~\cite{Fleischhauer00, Liu01, Novikova07, Gorshkov07}.

The rest of this paper is structured as follows.
In Section~\ref{sec:ocqt} we first review the Monte-Carlo wavefunction method
and the two preeminent gradient-based methods of quantum optimal control,
Gradient-Ascent-Pulse-Engineering (GRAPE) and Krotov's method.
Then, in Section~\ref{sec:trajoct}, we develop the central result of this paper,
a trajectory-based variant of Krotov's method.
We illustrate the method in a numerical case study in
Section~\ref{sec:casestudy}.
We first introduce the network model for a chain of cavities containing trapped
atoms.
Then, in Sections~\ref{sec:octresult}--\ref{sec:noise}, we present the result of
applying optimal control to the creation of a dark state in that system, and
analyze the convergence and the noise properties of the optimized pulses.
Section~\ref{sec:conclusion} concludes.

\section{Optimal Control via Quantum Trajectories}%
\label{sec:ocqt}

\subsection{The Monte-Carlo Wavefunction method}%
\label{sec:mcwf}

We consider the problem of initializing a quantum network from a well-defined
separable state $\Ket{\Psi(t=0)}$ to a (generally entangled) state
$\Ket{\Psi_{\mbox{\scriptsize tgt}}(t=T)}$.
The network is characterized by a Hamiltonian $\Op{H} = \Op{H}_0 + \sum_i u_i(t)
\Op{H}_i$ with one or more control fields $u_i(t)$, and a set of Lindblad
operators $\{\Op{L}_l\}$ to describe dissipative processes.
In the Markov limit~\cite{Breuer07}, the dynamical map $\Op{\rho}(t) =
\DynMap(t,0) \Op{\rho}(0)$ denotes the solution of the master equation
\begin{equation}
  \frac{\partial \Op{\rho}}{\partial t}
  = -\frac{i}{\hbar} \Lindbladian \Op{\rho}
  = -\frac{i}{\hbar}\left[\Op{H}(\{u_i(t)\}), \Op{\rho}\right] +
    \Lindbladian_D \Op{\rho}
  \label{eq:lindblad}
\end{equation}
with the boundary condition $\Op{\rho}_0 = \Ket{\Psi(0)}\!\Bra{\Psi(0)}$ and the
dissipator $\Lindbladian_D$.

For a network consisting of more than a trivially small number of nodes, in
general, the exponential growth of the Hilbert space will quickly render the
numerical treatment of the system dynamics in Liouville space unfeasible.
Significantly less resources are required to simulate the dynamics through
quantum trajectories.
In this case, $\Op\rho(t)$ is approximated as an average of the pure states
$\KetBra{\Psi_k(t)}{\Psi_k(t)}$ resulting from $M \rightarrow \infty$
trajectories indexed by $k$,
\begin{equation}
  \Op{\rho}(t)
  = \lim_{M\rightarrow\infty} \frac{1}{M} \sum_{k}
  \KetBra{\Psi_k(t)}{\Psi_k(t)}\,.
  \label{eq:rho_traj_expansion}
\end{equation}
Each trajectory $\Ket{\Psi_k(t)}$ is a possible, statistically independent,
pure-state realization of the system dynamics, taking into account dissipative
processes.
There are several ways to obtain the trajectories, corresponding to different
physical measurements on the environment (e.g.\
the field into which the system decays), and governed by different stochastic
equations of motion.
Measurements of one or both of the quadratures of the field result in a quantum
diffusion equation (QSDE)~\cite{PercivalQSDBook, Wiseman93}, while photon
counting results in quantum jumps (MCWF)~\cite{DumPRA1992,MolmerJOSA1993}.
For the purposes of evaluating the density matrix via~\Eq{rho_traj_expansion},
the various kinds of trajectories are equivalent.
We will focus on the quantum jump method because it is the most straightforward
to realize numerically.

The \emph{MCWF algorithm} for propagating the trajectory state $\Ket{\Psi_k}$
is~\cite{PlenioRMP1998}
\begin{algorithmic}[1]
  \State{}
    define the non-Hermitian effective Hamiltonian
    \begin{equation}
    \Op{H}_{\effective} = \Op{H} + \sum_l \Op{L}_l^{\dagger} \Op{L}_l
    \end{equation}
  \State{}
    draw a random number $r \in [0,1)$
  \State{}
    propagate until $\Abs{\Psi_k(t)}^2 = r$
  \State{}
    apply an instantaneous quantum jump
    \begin{equation*}
      \Ket{\Psi_k(t)}
      \rightarrow
        \frac{\Op{L}_l \Ket{\Psi_k(t)}}{\Norm{\Op{L}_l \Ket{\Psi_k(t)}}}\,,
    \end{equation*}
    choosing $\Op{L}_l$ from the set of all Lindblad operators with probability
    \begin{equation}
      \prob(\Op{L}_l)
      = \Braket{\Psi_k(t)|\Op{L}_l^{\dagger} \Op{L}_l | \Psi_k(t)}
    \end{equation}
  \State{}
    draw a new random number
    $r \in [0, 1)$ and continue the propagation
  \State{}
    normalize any resulting $\Ket{\Psi_k(t)}$.
\end{algorithmic}
In performing the propagation with discrete time-steps, as is essential
numerically, it is crucial to determine the jump times in step 3 with high
precision (through interpolation and bisection), and in this way multiple
instantaneous jumps can be included within one time step.
Using this approach, relatively long time steps are possible.
An alternative ``first-order'' scheme~\cite{MolmerJOSA1993} in which at most one
jump may happen in each time step is simpler to implement, but requires small
time steps to be accurate.

\subsection{Optimal Control with GRAPE and Krotov's method}%
\label{sec:oct}

We seek optimal solutions $u_i(t)$ to minimize the error functional
\begin{equation}
  J_T = 1 - \bigLBraket{\Op{\rho}(T)}{\Op{P}_{\mbox{\scriptsize tgt}}}\,,
  \label{eq:functional}
\end{equation}
for a fixed duration $T$, where $\Op{P}_{\mbox{\scriptsize tgt}}$ is the
projector onto the target state $\Ket{\Psi_{\mbox{\scriptsize tgt}}}$.
We use the notation $\LBraket{a}{b} \equiv \tr[a^\dagger b]$ for the
Hilbert-Schmidt overlap of two operators.

There are well-established gradient-based methods of numerical optimal control
to solve this problem.
The two preeminent ones are Gradient-Ascent-Pulse-Engineering
(GRAPE)~\cite{KhanejaJMR05}, usually as a quasi-Newton method in combination
with the L-BFGS-B algorithm~\cite{ByrdSJSC1995,ZhuATMS97}, and Krotov's
method~\cite{KrotovBook, PalaoPRA2003,ReichJCP12}.
Both of these are \emph{iterative}: They start from ``guess'' fields
$u_i^{(0)}(t)$ and calculate updates $\Delta u_i(t) = u_i^{(1)}(t) -
u_i^{(0)}(t)$ that decrease the value of the target functional; for the next
iteration, $u_i^{(1)}(t)$ becomes the guess, and the procedure repeats until
convergence is reached.

The GRAPE method starts with the assumption that $u_i(t)$ with $t \in [0, T]$ is
discretized into $n_t$ time steps, such that the time evolution up to a point
$t_j$ is
\begin{equation}
  \Op{\rho} (t_j)
  = \DynMap(t_j,0) \, \Op{\rho}(0)
  = \Biggl(\, \prod_{j'=j}^{1} \DynMap_{j'} \Biggr) \Op{\rho}(0)
\end{equation}
where $\DynMap_j \equiv \DynMap(t_j, t_{j-1})$ denotes the dynamical map
$\Op{\rho}(t_{j-1}) \rightarrow \Op{\rho}(t_{j})$, with $t_0 = 0$ and $t_{n_t} =
T$.
The indices in the product run backwards to account for time ordering.
The control fields are approximated as constant within the time interval,
$u_{ij} \equiv u_i([t_{j-1}, t_j])$.
Each time-local value $u_{ij}^{(0)}$ is modified according to the gradient,
\begin{equation}
  \Delta u_{ij}
   \propto \frac{\partial J_T}{\partial u_{ij}}
  = - \biggLBra{\Op{P}^{(0)}(t_j)}
     \frac{\partial \DynMap_j}{\partial u_{ij}}
     \biggLKet{\Op{\rho}^{(0)}(t_{j-1})} \,,
   \label{eq:gradient}
\end{equation}
where
\begin{equation}
  \Op{P}^{(0)}(t_j)
  = \Biggl(\, \prod_{j'=j+1}^{n_t} \DynMap_{j'}^{(0) \dagger} \Biggr)
    \Op{P}_{\mbox{\scriptsize tgt}}
  \label{eq:bwprop}
\end{equation}
is the target state backward-propagated with the conjugate Lindbladian.
The superscript zero indicates the dynamics under the guess controls.

Krotov's method takes a slightly different approach.
Starting from time-continuous guess controls $u_i^{(0)}(t)$, it analytically
constructs pulse updates $\Delta u_i(t)$ that minimize an
extended functional, commonly
\begin{equation}
  J = J_T + \sum_i
      \frac{\lambda_i}{S_i(t)}
      \int_{0}^{T} [
        \underbrace{u_i^{(1)}(t) - u_i^{(0)}(t)}_{= \Delta u_i(t)}]^{2}
      \dd t\,,
  \label{eq:krotov_functional}
\end{equation}
where $\lambda_i$ is an arbitrary weight determining the overall magnitude of
$\Delta u_i(t)$, and $S_i(t) \in [0, 1]$ is a shape function that may be used to
enforce boundary conditions.
As the optimization converges, $\Delta u_i(t) \rightarrow 0$, such that $J$ and
$J_T$ become equivalent.
The control problem is solved by the update equation
\begin{equation}
  \Delta u_i(t)
  = \frac{S_i(t)}{\lambda_i} \Im \biggLBra{\Op{P}^{(0)}(t)}
     \frac{\partial \Lindbladian}{\partial u_{i}(t)}
     \biggLKet{\Op{\rho}^{(1)}(t)} \,,
   \label{eq:krotovupdate}
\end{equation}
where $\Op{P}^{(0)}(t)$ is a state backward-propagated with the conjugate
Lindbladian, cf.~\Eq{bwprop}, with the boundary condition
\begin{equation}
  \Op{P}(T)
  = -  \frac{\partial J_T}{\partial \LBra{\rho(T)}}
  = \Op{P}_{\mbox{\scriptsize tgt}}\,,
  \label{eq:krotov_boundary}
\end{equation}
where we use a notation analogous to Ref.~\cite{ReichJCP12} for the derivative
with respect to the formal co-state in inner products in Liouville space.
Only now is the update $\Delta u_i(t)$ discretized to a time grid ($\Delta
u_i(t) \rightarrow \Delta u_{ij}$, $t \rightarrow t_j$), resulting in a formula
that superficially resembles~\Eq{gradient}, with at least two crucial
differences:

First, the forward-propagated state $\rho(t)$ uses the updated controls
$u_i^{(1)}(t)$ in~\Eq{krotovupdate}, and the guess controls $u_i^{(0)}(t)$
in~\Eq{gradient}.
This makes Krotov's method \emph{sequential} (updates at later times depend on
updates at earlier times), whereas GRAPE is \emph{concurrent} (updates are
independent).

Second, the local derivative term in each equation differs; The gradient
derivative $\partial \DynMap_{j} / \partial u_{ij}$ can be evaluated similarly
to $\DynMap_j$ itself, under the assumption that $\DynMap_{j}$ is a simple
exponential~\cite{FouquieresJMR2011}.
In comparison, $\partial \Lindbladian / \partial u_{ij}$ is more straightforward
to evaluate (and instantaneous). In full generality, $\Lindbladian$ in
\Eq{krotovupdate} is the right-hand side of the equation of motion for
$\Op{\rho}$, whether or not this is a master equation in Lindblad
form~\cite{ReichJCP12}.
For the particular form \Eq{lindblad} with a time-independent dissipator,
$\partial \Lindbladian / \partial u_{ij} = [\Op{H}_i, \rho(t_j)]$.

The backward-propagated state $\Op{P}^{(0)}(t)$ is the same for Krotov's method
and GRAPE only for the specific case of a direct state-to-state transfer,
functional~\Eq{functional}.
More generally, in Krotov's method, the boundary condition is determined
explicitly by~\Eq{krotov_boundary}, which for other optimization tasks (e.g.,
gate optimization) does not yield states identical to the target states (but
usually, a linear combination of them).
Even though for GRAPE, the boundary condition also implicitly depends on the
choice of functional, for all optimization tasks in quantum control that we are
aware of, the backwards propagation always starts from the target states.
While we have assumed linear controls and a convex functional, Krotov's method
can be also be extended to go beyond these assumptions~\cite{ReichJCP12}.

\subsection{Krotov's method for quantum trajectories}%
\label{sec:trajoct}

There are two possibilities for rewriting the optimization in terms of quantum
trajectories.
First, we can insert the expansion of $\Op{\rho}$ \Eq{rho_traj_expansion} into
the functional \Eq{functional} to find
\begin{equation}
  J_T = 1 - \lim_{M\rightarrow\infty} \frac{1}{M}\sum_{k=1}^{M}
          \big\vert
            \underbrace{%
                \Braket{\Psi_k(T) | \Psi_{\mbox{\scriptsize tgt}}}}_{%
              \equiv \tau_k}
          \big\vert^{2}\,.
  \label{eq:functional_traj}
\end{equation}
That is, the fidelity is simply the average of the fidelities from each
trajectory.
We seek a single control that simultaneously implements the state-to-state
transition for every $\ket{\Psi_k}$.

We now wish to ask, given the form of $J_T$, how the update rules for GRAPE and
Krotov's method might be written in terms of the individual trajectories.
For GRAPE this is not so simple; the form of the update rule is based on the
fact that the evolution of $\Op{\rho}$ in each time step is given by the
application of a linear operator $\DynMap_{j}$, and this is no-longer true for
the trajectories.
Second, in a given time step the evolution of a given trajectory may contain one
or more jumps at random times, which further complicates the calculation of the
derivative with respect to the controls $u_{ij}$.

In contrast, for Krotov's method the functional~\Eq{functional_traj} leads to
the update equation~\cite{PalaoPRA2003}
\begin{equation}
  \Delta u_i(t)
  = \frac{S_i(t)}{M\lambda_i} \sum_{k=1}^{M}
    \underbrace{\Im \Braket{\chi_k^{(0)}(t) | \Op{H}_i | \Psi_k^{(1)}(t)}}_{%
      \equiv \Delta u_{ik}(t)}\,,
    \label{eq:independent_traj_update}
\end{equation}
where $\chi_k^{(0)}(t)$ is backward-propagated with the boundary condition
\begin{equation}
  \chi_k^{(0)}(T)
  = -  \frac{\partial J_T}{\partial \Bra{\Psi_k}}
  =  \tau_k^{(0)} \Ket{\Psi_{\mbox{\scriptsize tgt}}}\,,
\end{equation}
where $\tau_k^{(0)}$ is $\tau_k$ from \Eq{functional_traj} evaluated for the
guess pulse.
The crucial difference is that for GRAPE, the gradient is evaluated for the time
evolution operator, whereas for Krotov's method, the gradient is with respect to
the equation of motion.
Only the latter is immediately compatible with the decomposition into quantum
trajectories.

From a numerical perspective, the update equation~\Eq{independent_traj_update}
can be evaluated with high efficiency: the contributions to the update $\Delta
u_{ik}(t)$ from every trajectory are completely independent.
If the numerical method is parallelized, so that different trajectories are
simulated on different processors, and a Message-Passing-Interface (MPI) is used
for interprocessor communication, only these scalar values need to be
communicated.
We thus reap the full benefit of the MCWF method: reduction of the overall
dimensionality by a factor of $d$, the dimension of the total Hilbert space.
The required memory per compute node is reduced directly by that factor; the
worst-case CPU time reduces by $d^3$ (as the fundamental operation in the time
propagation is a matrix-vector multiplication).

We can obtain an alternative formulation of the control problem in terms of
quantum trajectories by leaving the functional~\Eq{functional} in terms of the
density matrix, and instead expanding the update equation in terms of the
trajectories.
For GRAPE this approach has the same issues as before, whereas for Krotov's
method this results in the update equation
\begin{equation}
  \fl
    \Delta u_{i}(t)
    =
      \frac{S_i(t)}{M^2\lambda_i} \sum_{k, k'=1}^{M}
      \Im \Big[
        \Braket{\xi_k^{(0)}(t) | \Op{H}_i | \Psi_{k'}^{(1)}(t)}
        \Braket{\Psi^{(1)}_{k'}(t) | \xi_k^{(0)}(t)}
      \Big]\,,
  \label{eq:cross_traj_update}
\end{equation}
using the decomposition
\begin{equation}
  \Op{P}^{(0)}(t)
  = \lim_{M\rightarrow\infty}\frac{1}{M} \sum_{k=1}^{M}
    \KetBra{\xi_k(t)}{\xi_k(t)}
\end{equation}
for the backward-propagation with $\ket{\xi_k^{(0)}(T)} =
\ket{\Psi_{\mbox{\scriptsize tgt}}}$.
The above update now incorporates \emph{cross-trajectory} information, including
overlaps of the propagated states from each trajectory with those from every
other trajectory.
Compared to the full density matrix optimization, this alternative method
maintains the saving in CPU time and memory that comes from evolving the
trajectories independently, but does incur significantly more communication
overhead.

Both trajectory formulations of Krotov's method, the ``independent trajectory''
and ``cross-trajectory'' formulations, guarantee monotonic convergence for $M
\rightarrow \infty$.
However, the numerical efficiency rests on the assumption that $M$ can be kept
reasonably small, ideally smaller than $d$ (and certainly smaller than $d^3$).
For a significantly larger number of trajectories, the total numerical resources
will approach or even surpass those required for a full density matrix
propagation.
Note that even if there is no benefit in the \emph{total} numerical resources,
the memory \emph{per compute node} is still reduced by a factor of $d$.

For finite (small) $M$, there is no guarantee of monotonic convergence.
This is especially true because the stochastic jumps are different between
iterations, as well as between the forward and the backward-propagation within
the same iteration.
Moreover, we expect the instantaneous jumps in the trajectory to cause jumps
also in the control update.
These jumps will average out both with the number of trajectories and the number
of control iterations, but they may slow down convergence and may introduce
noise into the optimized control field.
To analyze how strongly the resulting optimized control fields are affected by
these two factors, and to what extent the cross-trajectory update equation might
be more robust, we now consider a practical example.

\section{Creation of an entangled dark state in a quantum network}
\label{sec:casestudy}

\subsection{Network Model}%
\label{sec:netmod}

\begin{figure*}[tb]
  \centering
  \includegraphics{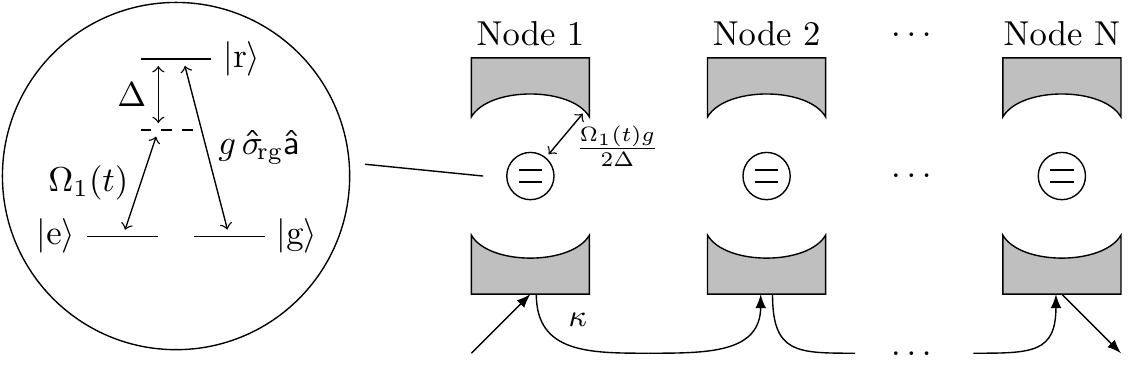}
  \caption{The structure of our network.
    A number of cavities, each containing a three-level atom in the Lambda
    configuration, are connected together by a traveling optical field that
    enters each cavity in turn via one of its end mirrors.
    Each cavity is coupled to the field via a circulator (not shown), so that
    the cavities only couple to field modes that are traveling from left to
    right.
    The field that is output from the first cavity thus enters the second, but
    not vice versa, in what is referred to as a ``cascade'' connection.}%
\label{fig:network}
\end{figure*}

We consider a network in which a number of ``nodes'' are connected via a
traveling-wave field, depicted in \Fig{network}.
The nodes are connected to the field via circulators so that the field
propagates from one node to the next in a single direction only, a configuration
often referred to as a ``cascade'' network.
Each node consists of a cavity containing a single atom.
The atom encodes a qubit in two degenerate states $\KetG$, $\KetE$, driven via a
Raman transition through an auxiliary level $\KetR$: $\KetG
\leftrightarrow\KetR$ couples through a time-dependent drive $\Omega(t)$ with
detuning $\Delta$, while $\KetR \leftrightarrow  \KetE$ is statically coupled
with coupling strength $g$, see \Fig{network}.
For large $\Delta$, the level $\KetR$ can be adiabatically
eliminated~\cite{CiracPRL1997}, resulting in an effective two-level system whose
coupling to the cavity is controlled via the drive $\Omega(t)$.

For a single node labeled $(i)$, the effective Hamiltonian in the rotating wave
approximation then consists of the drift term $\Op{H}_0$ and a driven
Jaynes-Cummings term $\Op{H}_{d}$.
With the cavity detuned from the driving field by $g^2/\Delta$ to compensate for
the Stark shift, we find
\begin{eqnarray}
  \Op{H}^{(i)} &=&
    \Op{H}^{(i)}_0 + \Op{H}^{(i)}_{d} \,,\\
  \Op{H}^{(i)}_0 &=&
    - \frac{g^{2}}{\Delta} \Op{a}_i^{\dagger} \Op{a}_i +
    \frac{g^{2}}{\Delta}
    \OpPiGround^{(i)} \otimes \Op{a}_i^{\dagger} \Op{a}_i\,,\\
  \Op{H}^{(i)}_{d} &=&
    - \frac{\ii \Omega_{i}(t) g}{2 \Delta}
    \left(\OpSigma{e}{g}^{(i)} \otimes \Op{a}_i -
          \text{c.c.}\right)\,,
\end{eqnarray}
where $\Op{a}_i$ is the cavity lowering operator, $\OpPiGround$ is the projector
on the qubit ground state, and $\OpSigma{e}{g}$ is the raising operator of the
qubit.
Leakage of photons out of the cavity is described by the Lindblad operator
\begin{equation}
  \Op{L}^{(i)} = \sqrt{2 \kappa} \, \Op{a}_i\,.
\end{equation}
There is also spontaneous decay from level $\ket{r}$ of the atom, but the decay
is suppressed for large $\Delta$ (the prerequisite for the adiabatic elimination
that yields the effective two-level system), and can be
neglected~\cite{CiracPRL1997}.

We now couple $N$ cavities to a traveling-wave field via circulators, so that
the field propagates from cavity to cavity in only one direction, as depicted in
\Fig{network}.
The resulting master equation describing the evolution of all the cavities is
obtained using input-output theory~\cite{gardiner85,Gardiner93,gardiner04}.
The explicit application of the theory is through the ``SLH'' formalism of Gough
and James~\cite{Combes17, GoughCMP2009, GoughITAC2009}.
While the calculation can be done by hand, it is tedious, and we instead use the
QNET software package, which automates the SLH formalism~\footnote{%
The python package QNET allows the user to construct quantum input-output
networks, and will calculate the master equation for any such
network~\cite{QNET}.
}.

The total Hamiltonian for the entire network is
\begin{eqnarray}
  \Op{H} = \sum_{i=1}^{N} \Op{H}^{(i)} + \sum_{i,j > i}^{N} \Op{H}^{(i,j)}\,,
  \label{eq:ham}\\
  \Op{H}^{(i,j)} =
  \ii \kappa \Op{a}_i^{\dagger} \Op{a}_j + \text{c.c.}
  \label{eq:ham_interaction}
\end{eqnarray}
Note that the connection via the field yields the additional static interaction
Hamiltonian $\Op{H}^{(i,j)}$ between \emph{all} nodes of the network (not merely
between nearest neighbors). The Lindblad operator for the entire network is
\begin{equation}
  \Op{L} = \sum_{i=1}^{N} \Op{L}^{(i)}\,.
\end{equation}
Thus, both the interaction between nodes and the total dissipation of the
network are proportional to the decay rate $\kappa$.
In order to create an entangled state quickly, $\kappa$ (the interaction) must
be large, which would seem to imply rapid decay of the network.
However, the form of the Lindblad operator allows for the individual summands
$\Op{L}_i$ to destructively interfere, canceling overall dissipation.
If the dark-state condition~\cite{Kraus08, Clark03, Cho11, Stannigel12}
\begin{equation}
  \forall t: \Braket{\Psi(t) | \Op{L}^{\dagger }\Op{L} | \Psi(t)} = 0
  \label{eq:darkstate}
\end{equation}
is fulfilled, an entangled state may be generated without loss of coherence.
Nonetheless, we must consider the dynamics of a quantum network to be inherently
highly dissipative, as the dark-state condition is only fulfilled for carefully
chosen control pulses that evolve the network exclusively within the dark-state
manifold.

In applying our numerical search method to the problem of generating an
entangled dark state in the network we assume that the field link between
cavities is the only significant source of loss.
Since the entangled dark state exists only because the same field link is
coupled to multiple systems, any sources of loss that are not similarly
collective will necessarily limit the fidelity with which the entangled state
can be prepared.
Such losses include spontaneous emission from the atomic excited state (which is
suppressed by the detuning), internal cavity losses, and loss \textit{en route}
between the cavities (as opposed to loss out of the open end of the field link,
which is the collective loss). These losses will be negligible so long as the
loss rates are small compared to the time it takes the protocol to prepare the
entangled state.
(This time scale is set by the rate at which the cavities decay out the open end
of the field link, which is the also the rate of the effective field-mediated
coupling between the cavities). We ignore all additional sources of loss in our
analysis because i) the effect of these loss sources cannot be reduced in any
significant way by optimizing the control protocol (since they do not possess
any associated dark states), and thus while this loss limits the achievable
fidelity it is not expected to change the convergence properties of the
numerical search method, and ii) to prepare high-fidelity entangled states
requires that any additional loss is small compared to that of the field link
itself.

\subsection{Optimized creation of an entangled dark state}%
\label{sec:octresult}

We assume that at time zero, the network is initialized in the state
$\Ket{\Psi(t=0)} = \Ket{eg\dots{}g}$.
That is, the atom in the cavity of the leftmost node is in the excited state,
while all other nodes (and all cavities) are in the ground state.
The state of the cavities, $\Ket{0\dots0}$ is implicit in our notation.
We now seek the control fields $\Omega_{i}(t)$ (one field for each node) that
will bring the system into an entangled dark state at a fixed final time $T$,
$\Ket{\Psi(t=T)} = \frac{1}{\sqrt{N}} \left( \Ket{eg\dots{}g} + \Ket{ge\dots{}g}
+ \cdots + \Ket{gg\dots{}e} \right)$.
The cavity for all nodes again must be in the ground state at time $T$.

The optimization problem is solved by minimizing~\Eq{functional} using Krotov's
method.
We first do this optimization using the full density matrix,
update~\Eq{krotovupdate}, and then -- for comparison -- using a trajectory
optimization, both for independent trajectories, functional~\Eq{functional_traj}
with update~\Eq{independent_traj_update}, and for the cross-trajectory method,
update~\Eq{cross_traj_update}.

One reason for choosing this particular physical system as an example to
illustrate our proposed optimization method is that the Hamiltonian~\Eq{ham}
preserves the total excitation in the system.
Thus, it is sufficient to model both the cavity and the atom at each node as a
two-level-system.
Moreover, when representing the state numerically, we need only include the
single-excitation subspace along with the ground state to account for the
possibility of spontaneous decay (the loss of the excitation out the open end of
the field that links the cavities). The effective dimension of the resulting
Hilbert space is $2 N + 1$, instead of the exponentially scaled $4^N$.
This makes a full density matrix optimization feasible even for relatively large
$N$, allowing us to compare against the trajectory optimizations.
We consider first the simplest case of $N=2$ nodes, with control fields
$\Omega_1(t)$ and $\Omega_2(t)$.
In this case, the dark state is the Bell state
\begin{equation}
  \Ket{\Psi}_{\mbox{\scriptsize tgt}} = \frac{1}{\sqrt{2}} \left( \Ket{eg} + \Ket{ge} \right)\,.
\end{equation}
This allows us to easily illustrate the characteristics of the optimized pulses
and node dynamics.
To ensure that our conclusions hold for larger networks, we then consider a
network of 20 nodes.
In principle, even larger networks would be numerically feasible.
However (as we will see), the optimization becomes increasingly harder, due to
the drastically larger search space resulting from $N$ independent control
pulses.
When the number of nodes is significantly higher that 20, we do not find good
control solutions within a reasonable number of iterations (even for the
optimization that employs the full density matrix). We expect that it might be
possible to sufficiently reduce the search space for larger networks by not
treating all pulses as fully independent.

For full generality, we use a dimensionless Hamiltonian where energies are
written in units of $g$.
Consequently, time can be expressed in units of $\hbar/g$.
For $\Delta = 100~g$ and $\kappa = g$, through systematical variation of $T$, we
find that the two-node dark state can comfortably be realized for $T=5~\hbar/g$.
For 20 nodes, we correspondingly increase the process duration to
$T=50~\hbar/g$.
In general, for a fixed number of nodes, the minimum time required to entangle
the nodes of the network is proportional to the interaction
Hamiltonian~\Eq{ham_interaction}, that is, proportional to the value of
$\kappa$.

\begin{figure*}[tb]
  \centering
  \includegraphics{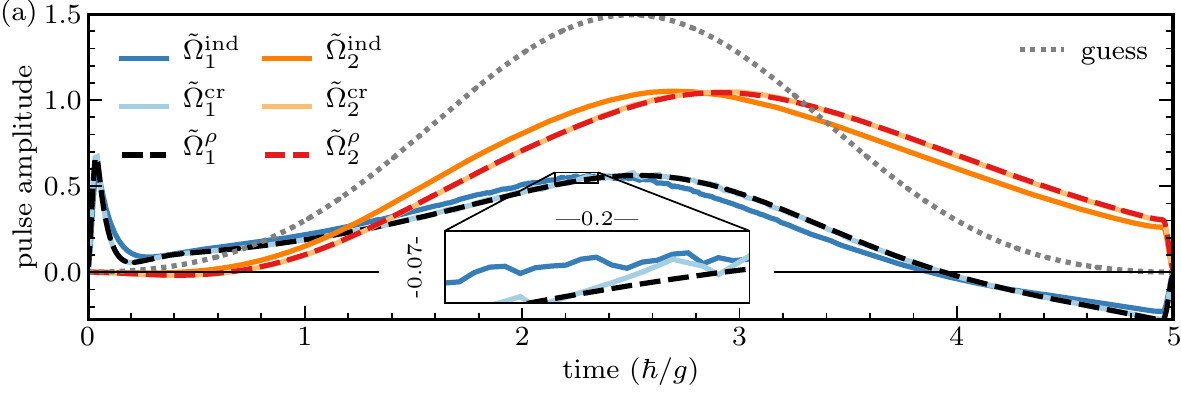} \\
  \includegraphics{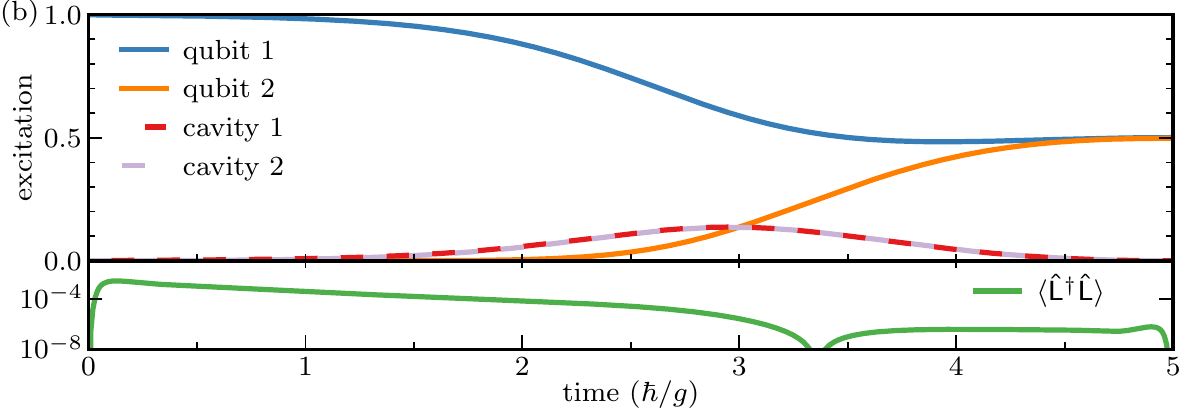}
  \caption{Optimized pulses and dynamics.
  (a) Optimized pulses for the creation of the dark state in a two-node
  network.
  $\Omega_{1,2}^{\rho}$ are the pulses for node 1, 2 resulting from optimizing
  using the density matrix.
  $\Omega_{1,2}^{\text{ind}}$ and $\Omega_{1,2}^{\text{cr}}$ are the pulses
  from optimizing using two independent trajectories, respectively two
  cross-referenced trajectories.
  All pulse amplitudes are normalized as $\tilde\Omega^{\rho, \text{tr}}_{1,2}
  \equiv \Omega^{\rho, \text{tr}}_{1,2} / 2 \Delta$.
  The guess pulse (identical for all nodes and optimizations)
  is indicated by the dotted gray curve.
  (b) Population dynamics under the optimized pulses $\Omega_{i}^{\rho}(t)$,
  as the expectation values of the qubit excitation, $\Avg{\OpPiExc^{(i)}}$,
  and the cavity excitation, $\Avg{\Op{a}^{\dagger}_{i}\Op{a}_{i}}$, with
  $i=1,2$.
  In the bottom, the value of the dark-state condition~\Eq{darkstate}.
  }
\label{fig:dynamics}
\end{figure*}

\Fig{dynamics} shows the result of an optimization for two nodes starting from a
simple Blackman-shaped guess pulse, see the dotted gray line in panel~(a). For a
direct density-matrix optimization using Krotov's method, that is the
minimization of~\Eq{functional} with update~\Eq{krotovupdate}, after 5000
iterations an error of $J_T = 1.3 \times 10^{-3}$ is achieved.
The optimized pulses $\Omega_1(t)$ and $\Omega_2(t)$ for the first and second
node are shown in \Fig[a]{dynamics}.
We show the pulses resulting from the full density matrix optimization
(superscript ``$\rho$''; black, red), from two independent trajectories
(superscript ``ind''; dark blue, dark orange), and from two cross-referenced
trajectories (superscript ``cr''; light blue, light orange). The corresponding
dynamics for the pulses resulting from the density matrix optimization are in
panel~(b), showing the smooth transition from the initial state $\ket{eg}$ to
the superposition of $\ket{eg}$ and $\ket{ge}$: the excitation of the first and
second qubit, the plotted expectation values $\Avg{\Pi_e^{(1)}}$ and
$\Avg{\Pi_e^{(2)}}$, correspond directly to the population of $\ket{eg}$ and
$\ket{ge}$, as the Hamiltonian preserves excitations ($\Braket{ee|ee} \equiv
0$).

As the interaction between the nodes is mediated by photons being emitted from
cavity 1 into cavity 2, both cavities are necessarily excited during the
transition (see the light and dark red dashed lines in \Fig[b]{dynamics}).
Nonetheless, the dynamics are (nearly) coherent, due to the system being in a
dark state (see the green line for $\Avg{\Op{L}^\dagger \Op{L}}$ in the bottom
of \Fig[b]{dynamics}, which is close to zero). The two cavities are
phase-shifted by $\pi$, that is $\Avg{\Op{a}_1} \approx - \Avg{\Op{a}_2}$,
resulting in destructive interference for the total dissipator.

As we have not taken into account any other dissipation channels (such as
spontaneous decay of the qubit), the deviation from the dark-state
condition~\Eq{darkstate}, is the only factor that limits the fidelity of the
transfer.
By continuing the optimization beyond 5000 iterations, the error could be made
arbitrarily small in principle.

Optimizing with independent trajectories, yields very similar results,
corresponding to the optimized pulses shown in \Fig[a]{dynamics} (dark orange,
dark blue) being close but not identical to the pulses obtained from direct
density matrix optimization (red, black). For the cross-trajectory method, the
dynamics and optimized pulses (light colored) are virtually indistinguishable
from the density matrix optimization.

In general, running both optimization methods with differing numbers of
trajectories, and the density matrix method, all yield comparable errors.
Nevertheless there are differences in the rate of convergence, as well as in
certain details of the pulse features.
As shown in the zoomed inset of \Fig[a]{dynamics}, there is some noise in the
optimized pulses obtained from trajectory optimization, compared to the
perfectly smooth result of the density matrix optimization.
This noise originates from the discontinuous jumps in the trajectories.
The optimized pulses and dynamics for the 20-node network (not shown) have the
same qualitative characteristics.
In the following, we will analyze how both the convergence rate and the noise in
the optimized pulses depend on the choice of optimization method and the number
of trajectories.

\subsection{Convergence of trajectory optimization}

\begin{figure*}[tb]
  \centering
  \includegraphics{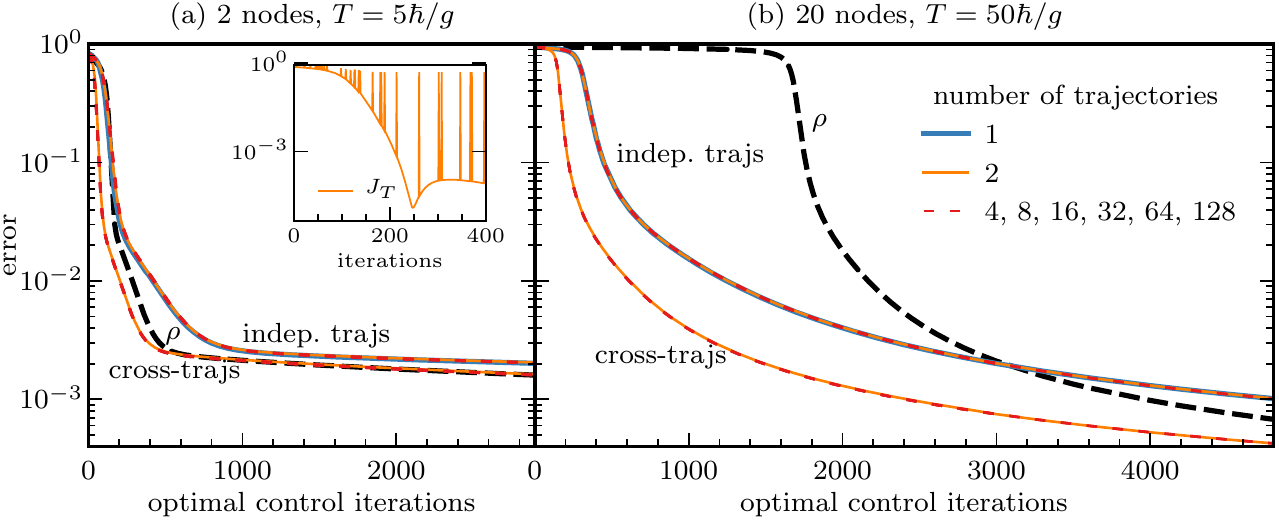}
  \caption{Comparison of convergence of optimization variants, for the two-node
  network, panel (a), and the 20-node network, panel (b).
  For using between 1 and 128 independent trajectories, the error
  $1 - \LBraket{\Op{\rho}(T)}{\Op{P}_{\mbox{\scriptsize tgt}}}$ after each iteration is shown.
  This compares to the optimization using between 2 and 128 cross-trajectories,
  and the optimization directly in Liouville space (dashed black curve).
  As the number of trajectories has almost no effect on the convergence and the
  curves are not distinguishable, all the curves for 4--128 trajectories are plotted with the
  same red dashed line style.
  All errors are evaluated from the exact density matrix evolution under the
  optimized pulses, which differs from the functional $J_T$ actually guiding the
  optimization.
  For the example of an optimization using two independent trajectories in the
  two-node network, the value of $J_T$ is shown in the inset of panel (a).
  }%
\label{fig:convergence}
\end{figure*}

\Fig{convergence} shows the convergence, as measured by the
error~\Eq{functional}, over the number of iterations in the control procedure.
In each iteration, the pulse is updated according to the appropriate update
equation.
Panel (a) shows the results for the two-node network, while panel (b) shows the
result for the 20-node network.

The convergence for the full density matrix optimization,
update~\Eq{krotovupdate}, is shown as the black dashed line.
It compares to the convergence for using 1-128 independent trajectories
(``indep.\ trajs''), update~\Eq{independent_traj_update}, and  2-128
cross-trajectories (``cross-trajs''), update~\Eq{cross_traj_update}.

We first observe that the optimization for the 20-node network is \emph{much}
harder than the optimization for 2 nodes, requiring several thousand iterations
to reach errors below $10^{-2}$.
In particular, the density matrix optimization has a long initial plateau before
starting to converge.
This behavior is exacerbated for more nodes, which is why we have limited
ourself to considering at most a 20 node network.
Remarkably, the use of a trajectory optimization performs significantly better
in this initial phase, largely avoiding the plateau.
We presume this is because the dissipative jumps affect the trajectories much
more strongly than the smooth decay affects the density matrix, leading to a
steeper gradient.
Both the two-node network and the 20-node network have the same qualitative
convergence characteristics; the larger number of nodes simply ``stretches''
them out.

The convergence of the cross-trajectory optimization consistently outperforms
the optimization using independent trajectories.
Asymptotically, the independent trajectories eventually lag behind the full
density matrix optimization, but only slightly so.
The cross-trajectory optimization asymptotically reaches the same error as the
full density matrix optimization (clearly for the two-node network, but we may
extrapolate this claim to hold also for 20 nodes).

A second remarkable observation is that the number of trajectories has almost no
effect on the convergence.
Thus, even a single trajectory, or two cross-referenced trajectories, can yield
a satisfactory result, a significant reduction of numerical effort compared to
the use of full density matrices.
It is worth pointing out that as long as the number of trajectories does not
exceed the number of CPU cores available on a compute node, it would be easy to
parallelize the method not using a message-passing interface, but instead using
a shared-memory approach (e.g.\ OpenMP). In this case, the additional numerical
overhead of the cross-trajectory optimization relative to independent
trajectories largely disappears.

For the trajectory optimizations, the error~\Eq{krotovupdate} that is shown in
\Fig{convergence} is not identical to the functional $J_T$ that the optimization
directly aims to minimize~\Eq{functional_traj}.
Because of the finite number of trajectories and the randomness inherent in the
quantum jumps, $J_T$ itself does not converge monotonically.
An example for two independent trajectories is shown in the inset of
\Fig{convergence}.
The loss of a photon from the cavity out of the open end of the field link (that
is, a quantum jump) in a given iteration results in a total loss (error 1.0), as
visible in the spikes in $J_T$.
In addition, the lower envelope ($J_T$ in the absence of photon loss) shows a
local minimum around 250 iterations, owing to the two competing requirements of
the optimization: achieving the desired target state (in the absence of photon
loss), and minimizing decay (the dark-state condition). For cross-trajectory
optimization, $J_T$ behaves similarly (not shown). The non-monotonic convergence
of $J_T$ should not be held against the trajectory optimization method: as shown
in \Fig{convergence} the actual physical error does decrease monotonically.

\subsection{Jump noise}%
\label{sec:noise}

\begin{figure*}[tb]
  \centering
  \includegraphics{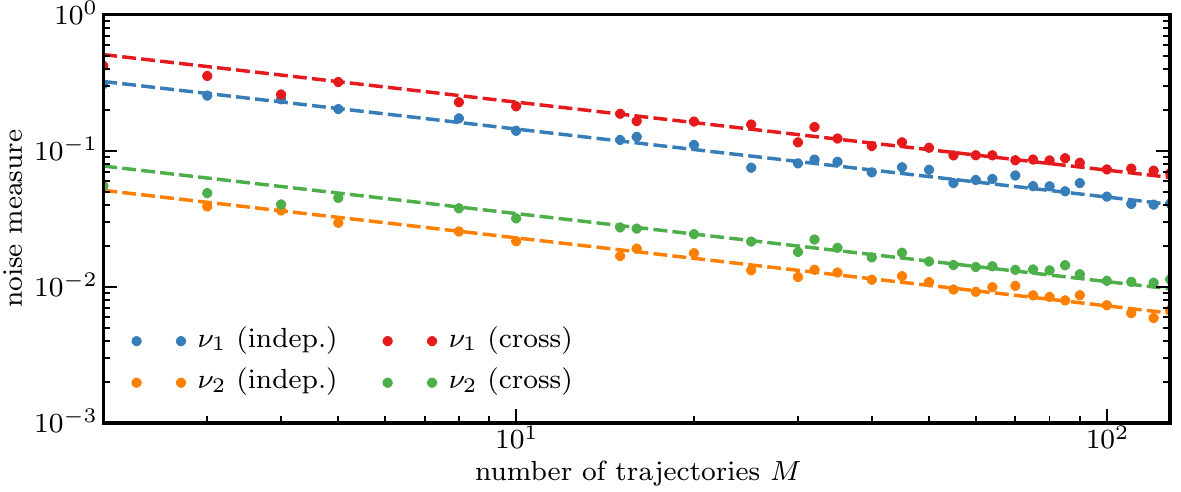}
  \caption{Scaling of noise artifacts in optimized pulses with the number of
  trajectories $M$. For both independent trajectories and cross-trajectory
  optimization, the noise measure according to~\Eq{noise} is shown for
  the pulse in each node. The dashed lines represent a fit to $\nu_i \propto
  1/\sqrt{M}$}%
\label{fig:noise}
\end{figure*}

The use of trajectory optimization results in some noise in the control fields,
cf.\ the inset in \Fig[a]{dynamics}.
To analyze how the magnitude of this noise scales with the number of
trajectories, we quantify the noise by comparing the optimized control field
$\Omega_i(t)$ to a smoothed pulse $\Omega_i^{\mbox{\scriptsize smooth}}(t)$,
\begin{equation}
  \nu_i = \int_{0}^{T} \Abs{\Omega_i(t) - \Omega_i^{\mbox{\scriptsize smooth}}(t)} \dd t\,.
  \label{eq:noise}
\end{equation}
For the purpose of this analysis, a good noise filter should generate an
$\Omega_i^{\mbox{\scriptsize smooth}}(t)$ that deviates from $\Omega_i(t)$ as
little as possible, while achieving a well-defined degree of smoothness.
A standard Savitzky-Golay filter~\cite{SavitzkyAC1964} is well-suited to solve
this problem~\cite{NumRecipesFortran}.

We find that the noise as defined in~\Eq{noise} asymptotically scales with the
number of trajectories $M$ as $\nu_i \propto 1/\sqrt{M}$.
This is illustrated in \Fig{noise}, which compares the noise measure for both
pulses of the two-node network, and for independent- and cross-trajectory
optimization, as a function of the number of trajectories $M$.
The results shown here use a five-point cubic convolute in the Savitzky-Golay
filter; this choice of window length and polynomial order only affects the
proportionality factor, i.e.\ the y-intercept in \Fig{noise}.

For independent trajectories, the $1/\sqrt{M}$ scaling holds for all values of
$M$.
The pulse $\Omega_1$ is noisier than the pulse $\Omega_2$, which is due to the
fact that $\Omega_2$ is closer to the original guess pulse than $\Omega_1$,
see~\Fig[a]{dynamics} -- any noise induced by a jump is proportional to the
magnitude of the pulse update in a given iteration.

The noise for the cross-trajectory optimization is systematically slightly
higher than that for independent trajectories.
We may conjecture that this is because the update~\Eq{cross_traj_update} is
quadratic in the (discontinuous) states, while the
update~\Eq{independent_traj_update} is only linear.
However, for a small number of cross-trajectories ($\lessapprox 5$), the noise
is smaller than the asymptotic $1/\sqrt{M}$ scaling, closer to the independent
trajectories.

In any case, for this specific example, the noise has little effect on the
fidelity, as we can see from \Fig{convergence}: the optimization result is
robust with respect to the level of noise introduced by the numerical procedure.
If this were not that case, and if smooth pulses are required, one may also
incorporate the smoothing directly into the optimization scheme by applying a
smoothing filter, or fitting to a parametrized curve, after each iteration (or
after some fixed number of iterations). The pulses resulting from the
optimization in the 20-node network (not shown) have noise characteristics
compatible with the above discussion.

\section{Conclusion}
\label{sec:conclusion}

We have formulated two variants of Krotov's method of optimal control based on
Monte-Carlo quantum jump trajectories for implementing state-to-state
transitions.
The fact that it is possible to apply Krotov's method directly to stochastic
trajectories is due to the fact that, unlike other methods, Krotov's method does
not require the calculation of the derivative of the propagator (which for
stochastic trajectories does not exist). The first variant considers $M$
independent trajectories, and optimizes the target in each trajectory, averaging
pulse updates from the trajectories.
It is easily parallelized to $M$ processes communicating with a message passing
interface (MPI). The second ``cross-trajectory'' variant obtains an update by
cross-referencing $M$ trajectories into a coarse approximation of the density
matrix evolution.
When parallelized using MPI, this produces considerable overhead.
However, for small $M$, a shared-memory parallelization model is efficient.

In any case, the use of trajectories provides a considerable numerical advantage
for systems that are strongly dissipative, but for which simulating the full
density matrix is not feasible due to the large dimension of the Hilbert space.
In many cases, the memory to store a density matrix of size $d^2$ is not
available, while a pure state of dimension $d$ can easily be stored and
propagated.
This is typically the case for quantum networks, and we have illustrated the
method for a simple example of creating an entangled dark state for a network of
20 cascaded cavities in which each contains a trapped atom.

We find that the use of trajectories in the optimization yields results
comparable to the full density matrix optimization.
In the case of cross-trajectory optimization, the convergence of the
optimization even outperforms the density matrix optimization.

Remarkably, the number of trajectories has almost no influence in the
convergence or the error that is achieved.
The number of trajectories does have an influence in the amount of noise
artifacts in the optimized pulses that result from the discontinuous (jump)
dynamics.
These scale as $1/\sqrt{M}$ in the number of trajectories.
At the same time, they have an only negligible effect on the dynamics and the
resulting error, and could also be removed by applying a smoothing filter to the
optimized pulses.

Thus, we can conclude that a relatively small number of trajectories (even just
a single trajectory) is sufficient to optimize a state-to-state transition in a
typical dissipative system.
Further, this trajectory approach is especially efficient when the optimal
protocol evolves the system within a dark (dissipationless) subspace, a
situation which arises, for example, both in quantum networks and atomic
ensembles~\cite{Fleischhauer00}.
The method naturally extends to optimization problems beyond a simple state
preparation, e.g.\ the realization of quantum gates~\cite{GoerzNJP2014}, or any
other optimization that is easily expressed in terms of the density matrix.
We have presented our optimization method and the example of the creation of a
dark state in the context of the master equation in Lindblad form.
However, the method extends to non-Markovian dynamics as well.
It has been shown that Krotov's method can be applied beyond the Markov
approximation and leads to an update equation that is structurally identical to
\Eq{krotovupdate}~\cite{OhtsukiJCP2003, HwangPRA2012, TaiPRA2014}, which can
then be rewritten in terms of quantum trajectories.
Also, instead of a quantum jump propagation, another trajectory method such as
quantum state diffusion~\cite{PercivalQSDBook} could be used.
In all cases, the feasibility of the optimization using Krotov's method results
from the fact that no analytic derivative of the inherently non-analytical
propagator is required.

\section*{Acknowledgments} 
The network model used in this paper was derived using the Python QNET
package~\cite{QNET}.
The propagation and optimization were done using the Fortran QDYN
library~\cite{QDYN}.
The pulse smoothing was realized through SciPy's signal processing
toolbox~\cite{scipy}.
This research was supported in part by ASD(R\&E) under their Quantum Science and
Engineering Program (QSEP), and by the Army High Performance Computing Research
Center (AHPCRC) (sponsored by the U.S. Army Research Laboratory under contract
No.
W911NF-07-2-0027).


\section*{References} 
\bibliography{trajoct}

\end{document}